%% file: liprior.tex
\documentclass[a4paper,11pt]{article}
\pdfoutput=1
\usepackage{jheppub}
\usepackage[T1]{fontenc}
\usepackage{graphicx}
\usepackage{amsmath}
\usepackage{amsfonts}
\usepackage{xspace}
\usepackage{subfig}
\usepackage{xcolor}
\usepackage{mathtools}
\usepackage{braket}
\usepackage{multirow}
\usepackage[utf8]{inputenc}
\usepackage{bm}
\usepackage{tikz}
\usepackage[compat=1.1.0]{tikz-feynman}
\usepackage{tgheros}
\usepackage{varwidth}
\usepackage{setspace}
\usepackage{float}
\usepackage{wrapfig}
\usepackage{url}

\makeatletter
\g@addto@macro\bfseries{\boldmath}
\makeatother

\PassOptionsToPackage{unicode}{hyperref}
\PassOptionsToPackage{naturalnames}{hyperref}

\usetikzlibrary{shapes,arrows,positioning,fit,calc,backgrounds}

\graphicspath{{./figures/}}

\tikzstyle{decision} = [diamond, draw, fill=blue!20, text badly centered, node distance=3cm, inner sep=2pt]
\tikzstyle{block} = [rectangle, draw, fill=blue!20, text centered, rounded corners, minimum height=2.5em]
\tikzstyle{line} = [draw, -latex']
\tikzstyle{cloud} = [draw, rectangle,fill=red!20, text width = 9em,
    minimum height=4em, rounded corners, node distance=2em]
\tikzstyle{bigbox} = [draw=black!50, thick, fill=blue!10, rectangle]

\tikzset{fit margins/.style={/tikz/afit/.cd,#1,
    /tikz/.cd,
    inner xsep=\pgfkeysvalueof{/tikz/afit/left}+\pgfkeysvalueof{/tikz/afit/right},
    inner ysep=\pgfkeysvalueof{/tikz/afit/top}+\pgfkeysvalueof{/tikz/afit/bottom},
    xshift=-\pgfkeysvalueof{/tikz/afit/left}+\pgfkeysvalueof{/tikz/afit/right},
    yshift=-\pgfkeysvalueof{/tikz/afit/bottom}+\pgfkeysvalueof{/tikz/afit/top}},
    afit/.cd,left/.initial=4pt,right/.initial=4pt,bottom/.initial=4pt,top/.initial=4pt}
    
\pgfdeclarelayer{back1}
\pgfdeclarelayer{back2}
\pgfdeclarelayer{back3}
\pgfdeclarelayer{back4}
\pgfsetlayers{back1,back2,back3,back4,main}

\title{Least-Informative Priors for $0\nu\beta\beta$ Decay Searches}
\author[a]{Frank F. Deppisch}
\author[a]{Graham Van Goffrier}

\affiliation[a]{Department of Physics and Astronomy, University College London, \\Gower Street, London WC1E 6BT, UK}

\emailAdd{f.deppisch@ucl.ac.uk}
\emailAdd{graham.vangoffrier.19@ucl.ac.uk}

\abstract{Bayesian parameter inference techniques require a choice of prior distribution which can strongly impact the statistical conclusions drawn. We discuss the construction of least-informative priors for neutrinoless double beta decay searches. Such priors attempt to be objective by maximizing the information gain from an experimental setup. In a parametrization using the lightest neutrino mass $m_l$ and an effective Majorana phase parameter $\Phi$, we construct such a prior using two different approaches and compare them with the standard flat and logarithmic priors in $m_l$.}

\keywords{Bayesian Inference, Least-Informative Prior, Maximum Entropy Prior, Neutrino Mass, Lepton Number Violation, Neutrinoless Double Beta Decay}

\begin{document}
\maketitle
\flushbottom
\newpage

\section{Introduction}
\label{sec:introduction}
\input{introduction}

\section{Neutrinoless Double Beta Decay and Neutrino Parameter Inference}
\label{sec:dbd}
\input{dbd}

\section{Least-Informative Priors}
\label{sec:lip}
\input{lip}

\section{Results}
\label{sec:results}
\input{results}

\section{Conclusion}
\label{sec:conclusion}
\input{conclusion}

\acknowledgments The authors would like to thank Matteo Agostini for useful discussion. The authors acknowledge support from the UK Science and Technology Facilities Council (STFC) via the Consolidated Grants ST/P00072X/1 and ST/T000880/1. G. V. G. also acknowledges support from the UCL Centre for Doctoral Training in Data Intensive Science funded by STFC, and from the UCL Overseas Research Scholarship / Graduate Research Scholarship.

\newpage
\appendix
\section{LIP Algorithm Flowchart}
\label{sec:flowchart}
\input{flowchart}
\clearpage

\bibliographystyle{JHEP}
\bibliography{liprior}
\end{document}

%% file: introduction.tex
Neutrinoless double beta ($0\nu\beta\beta$) decay is a hypothetical process of crucial interest due to its sensitivity both to the neutrino mass scale and to lepton-number violation. Direct searches for the decay, alongside neutrino oscillation studies and other probes of neutrino masses such as cosmological bounds \cite{Planck2018} and single-beta decay measurements \cite{KATRIN2015}, are key to the improvement of our understanding of neutrinos. While a measurement of the $0\nu\beta\beta$ decay rate has not yet been made, upper bounds have been placed on the effective $0\nu\beta\beta$ mass $m_{\beta\beta}$, from which constraints on the neutrino mass scale and Majorana phases may be inferred.

Lacking evidence to the contrary, early formulations of the Standard Model (SM) took neutrinos to be massless. However, data suggesting the occurrence of neutrino flavour oscillations has accumulated since the 1960s, beginning with the Homestake Experiment \cite{Cleveland1998} and culminating with the combined observation of oscillations for atmospheric neutrinos by Super-Kamiokande (SK), for solar neutrinos by the Sudbury Neutrino Observatory (SNO) \cite{Ahmad2001} and for reactor antineutrinos by the KamLAND-Zen Experiment \cite{Kamland2003}. From this it is clear that oscillations necessitate a nonzero mixing angle as well as a nonzero mass difference, i.e. no more than one neutrino mass may be zero. For a model describing the three flavours $\nu_e$, $\nu_\mu$, $\nu_\tau$ of neutrinos in the SM, mixing between weak and mass eigenstates is described by the Pontecorvo–Maki–Nakagawa–Sakata (PMNS) matrix, and again nonzero mixing angles and two nonzero mass eigenvalues are required, with the additional possibility of $CP$ violation.

The realisation that at least two generations of neutrinos are massive leads to natural questions: What is the mechanism responsible for the neutrino masses, and how can all three masses be measured? Oscillation itself lends evidence to the latter question, as measurements have been made of all PMNS matrix elements as well as mass-squared differences $\Delta m_{21}^2 \sim 7.4\cdot 10^{-5}$~eV$^2$ and $|\Delta m_{31}^2| \sim 2.5\cdot 10^{-3}$~eV$^2$ \cite{Nufit2018}~\footnote{A more recent global fit of oscillation parameters, which does not significantly impact on our results, may be found in \cite{deSalas2020}.}. The sign of $\Delta m_{21}^2$ is known, due to solar matter effects on oscillation \cite{deSalas2017}, leading to two candidate mass hierarchies: the normal-ordering (NO) $m_1 < m_2 < m_3$, and the inverse-ordering (IO) $m_3 < m_1 < m_2$. Throughout this report, we take these oscillation parameters to have their best-fit values from NuFIT 4.0 + SK \cite{Esteban2019}, for which uncorrelated errors are also accounted for in simulation.

Further information on neutrino parameters is provided by cosmological observations, as massive neutrinos play the unique role of both radiation during the early baryon acoustic oscillation epoch, and hot dark matter during later formation of large-scale structure. With the reasonable assumption of equal number densities for all flavours, the sum of masses $\sum m_i$ is proportional to the total energy density of neutrinos in non-relativistic eras. As a result, this quantity is observable in the redshift fluctuations of the cosmic microwave background, which are controlled by energy density at last-scattering via the integrated Sachs–Wolfe effect, as well as in a suppression of structural matter fluctuations due to free-streaming neutrinos. The latest fits from Planck observatory data \cite{Planck2018} place an upper bound $\sum m_i < 0.12$~eV at $95\%$ confidence, and efforts in this direction have great potential for further precision.

Our focus is on $0\nu\beta\beta$ decay, which is a hypothetical nuclear transition $2n\to 2p^+ + 2e^-$ \cite{Deppisch:2012nb, Dolinski:2019nrj}. Total electron number is violated by two units and the process is therefore not permitted in the SM with zero neutrino masses. It instead proceeds if the SM Lagrangian is extended by a Majorana mass term of the form $-\frac{1}{2}m\overline{\nu_L^C}\nu_L$. The decay process is sensitive to Majorana neutrinos with an observable effective mass
\begin{align}
\label{eq:effmass}
    m_{\beta\beta} = \sum_{i=1}^3 U_{ei}^2 m_i,
\end{align}
where PMNS matrix $U$ includes Majorana phases that are unobservable in oscillation experiments. 

The focus of the present work is the development of computational techniques for data-driven Bayesian inference on the $0\nu\beta\beta$ parameter space. Bayesian methodologies such as Markov Chain Monte Carlo (MCMC) require a choice of prior distribution, which can strongly influence derived bounds. While it is standard to apply flat priors to bounded parameters such as Majorana phases, for unbounded parameters such as neutrino masses there is less of a consensus. Recent Bayesian analyses have either preferred log-flat priors for their scale-invariance \cite{Agostini2017}, or have considered both flat and log-flat priors \cite{Caldwell2017} in order to demonstrate the strong dependence of quantities such as discovery probability on prior selection. In this paper we examine a class of least-informative priors (LIPs) and apply them to the analysis of $0\nu\beta\beta$ decay. LIPs are constructed by numerical maximisation of the expected Kullback-Leibler divergence between posterior and prior distributions \cite{Bernardo1979}, which is taken to represent inferential information gain and therefore to indicate a minimum quantity of information contained in the prior distribution.

In the frequentist interpretation of statistics, a perfect theory is believed to exist, whose parameters have an unknown but fixed value. Bayesian statistics, however, treats theories as having associated degrees of belief. Provided some data, this approach allows a practitioner to infer a sensible probability that a candidate theory is correct. Given that the primary goal of particle physics experimentation is to reject or accept candidate theories and refine their parameters, this Bayesian inference methodology is a powerful analysis tool. We consider the light neutrino exchange mechanism for $0\nu\beta\beta$ decay, and use published or Poisson-estimated likelihood functions from cosmological observations of $\sum m_i$ and direct searches for $m_{\beta\beta}$ to derive bounds on the neutrino masses and Majorana phases.

In Section~\ref{sec:dbd}, we briefly summarize the key aspects of neutrinoless double beta decay and how MCMC studies are being used to infer neutrino parameters. We also introduce an effective parameter which combines the effect of the Majorana phases. In Section~\ref{sec:lip}, we outline the theoretical foundations for least-informative priors and we detail an algorithm for generating LIPs applicable to our physics context. Our results are presented in Section~\ref{sec:results}. Conclusions and an outlook are featured in Section~\ref{sec:conclusion}.

%% file: dbd.tex
Thinking of the three terms in Eq.~\eqref{eq:effmass} as complex numbers, two relative phases $-\pi \leq \alpha, \beta < \pi$ are sufficient to describe $m_{\beta\beta}$. Explicitly writing out the PMNS mixing matrix elements, this yields
\begin{align}
	m_{\beta\beta} 
	= c^2_{12}c^2_{13} m_1 + s^2_{12}c^2_{13} m_2 e^{i\alpha} 
	+ s^2_{13} m_3 e^{i\beta} 
	= A + B e^{i\alpha} + C e^{i\beta},
	\label{eq:mbb}
\end{align}
where $s^2_{ij}$ and $c^2_{ij}$ are shorthand for $\sin^2\theta_{ij}$ and $\cos^2\theta_{ij}$, respectively. The real-valued coefficients $A$, $B$ and $C$ are functions of the neutrino masses and mixing parameters. The neutrino masses are not independent of each other, with the mass-squared splitting values fixed by oscillations, where the lightest mass $m_l$ can be chosen as a free parameter; $m_l = m_1$ if the neutrinos are normally ordered (NO) and $m_l = m_3$ if they are inversely ordered (IO). $0\nu\beta\beta$ experiments measure the  half-life $T_{1/2}({}^A\text{X})$ of the decay of an isotope ${}^A$X \cite{Agostini2017}, which is connected to the effective double beta mass as
\begin{equation}
	T^{-1}_{1/2}({}^A\text{X}) = |m_{\beta\beta}|^2 G_{0\nu} |\mathcal{M}_{0\nu}|^2.
\end{equation}
Here, $G_{0\nu}$ is a phase-space factor encoding the leptonic part of the process including Coulomb effects between the nucleus and outgoing electrons, and $\mathcal{M}_{0\nu}$ is the nuclear matrix element (NME) of the underlying nuclear transition. Both depend on the isotope in question. Due to the large size of the relevant nuclei considered and correlations between nucleon states, the $0\nu\beta\beta$ NMEs are challenging to calculate, and disagreement between different nuclear models is a significant source of theoretical error in $0\nu\beta\beta$ studies \cite{Pas2015}. A large number of experiments have succeeded in placing increasingly restrictive lower bounds on the $0\nu\beta\beta$ decay half-life. Recently, the KamLAND-Zen and GERDA experiments have determined the limits $T_{1/2}({}^{136}\text{Xe}) > 1.1\cdot 10^{26}$~yr \cite{Kamland2016} and $T_{1/2}({}^{76}\text{Ge}) > 1.8\cdot 10^{26}$~yr \cite{GERDA2020} at a $90\%$ confidence level \footnote{A Bayesian analysis is also performed by the GERDA collaboration, obtaining a slightly weaker bound of $T_{1/2}({}^{76}\text{Ge}) > 1.4\cdot 10^{26}$~yr for the same experimental data \cite{GERDA2020}.}.

\subsection{An Effective Majorana Phase Parameter}

\begin{figure}[t!]
	\centering
	\includegraphics[width=9.3cm,trim={8cm 1cm 8cm 1cm},keepaspectratio]{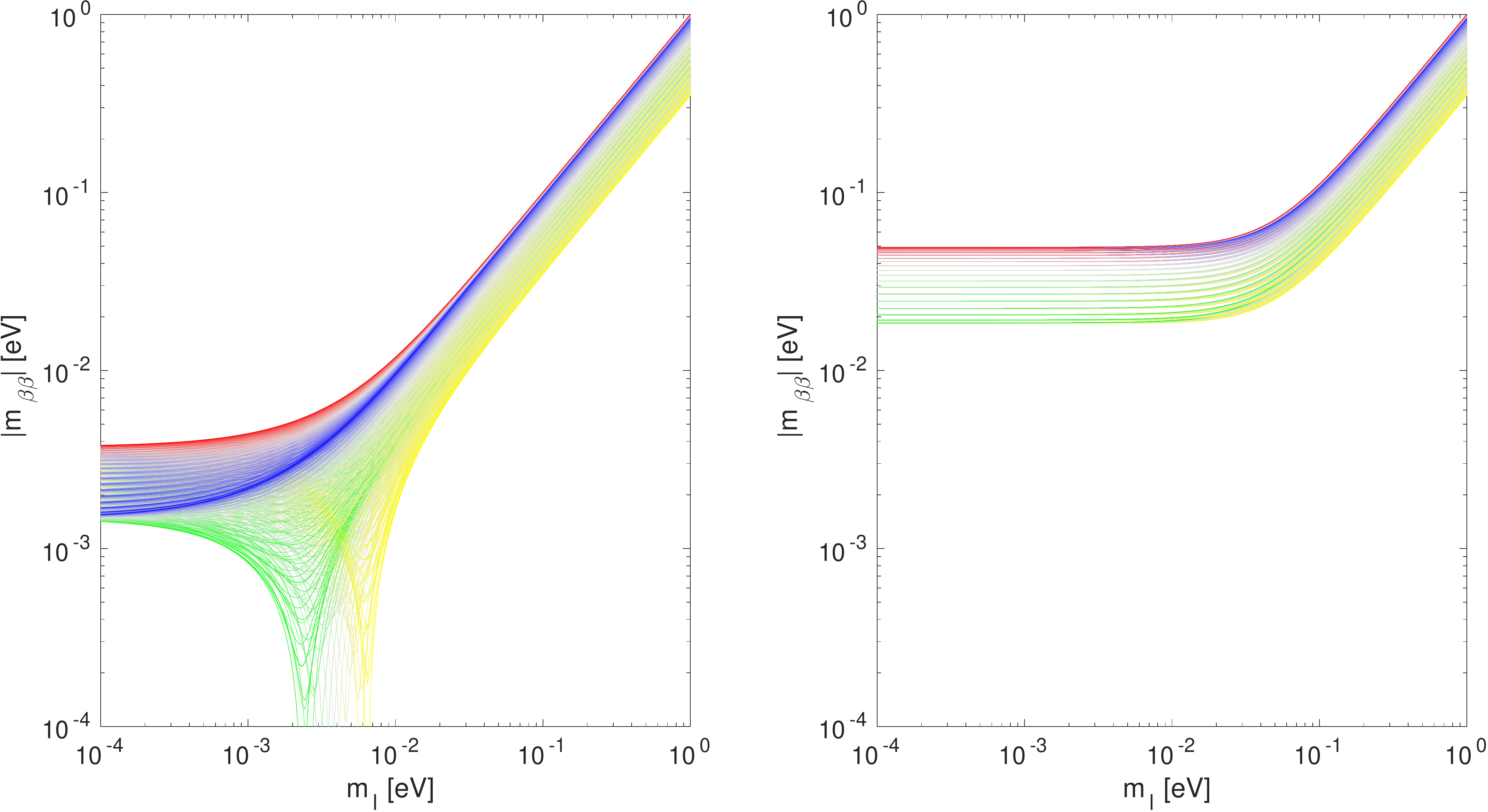}
	\caption{Effective mass $|m_{\beta\beta}|$ as function of the lightest neutrino mass $m_l$ for constant sets of Majorana phases $(\alpha,\beta)$, in the NO case (left) and IO case (right). The trajectories are colour-coded, interpolating (toroidally) between the four corners $(\alpha,\beta) = (0,0)$ (red), $(0,\pi)$ (blue), $(\pi,0)$ (green) and $(\pi,\pi)$ (yellow).}
	\label{fig:flat2phase}
\end{figure}
The parametrization of $0\nu\beta\beta$ decay arising from the neutrino mass matrix, while physically natural, introduces unnecessary complications into our inference. The Majorana phases entering it cannot be determined individually \cite{Deppisch:2004kn} and a combination of both leads to the band structure observed. Fig.~\ref{fig:flat2phase} shows a continuum of $|m_{\beta\beta}|$ trajectories against the lightest neutrino mass $m_l$ for constant choices of the Majorana phases $(\alpha,\beta)$, where both $\alpha$ and $\beta$ are uniformly scanned over the range $[-\pi,\pi]$. In a 2D rectilinear uniform distribution, volume effects imply that a point is more likely than not to be found near the boundary of the rectangle, and in particular near the corners. For both types of ordering, this and the functional dependence result in four bands of markedly-high density, which are often but not always located near the theoretical limits on $|m_{\beta\beta}|$.

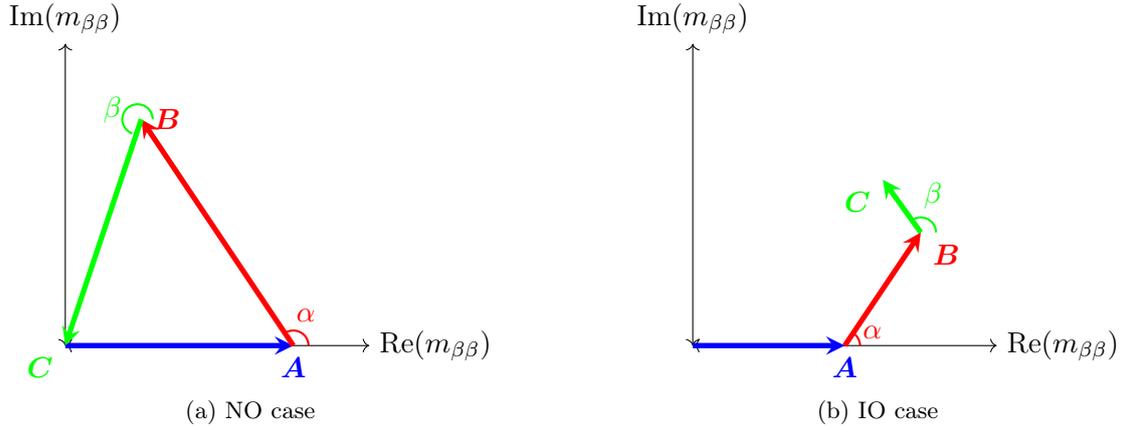
\begin{figure}[t!]
	\centering
	\subfloat[NO case]{%
		\begin{tikzpicture}
			\draw[thin,gray!40] (0,4) grid (0,4);
			\draw[<->] (0,0)--(4,0) node[right]{Re$(m_{\beta\beta})$};
			\draw[<->] (0,0)--(0,4) node[above]{Im$(m_{\beta\beta})$};
			\draw[line width=2pt,blue,-stealth](0,0)--(3,0) node[anchor=north]{$\boldsymbol{A}$};
			\draw[line width=2pt,red,-stealth](3,0)--(1,3) node[anchor=west]{$\boldsymbol{B}$};
			\draw[line width=2pt,green,-stealth](1,3)--(0,0) node[anchor=north east]{$\boldsymbol{C}$};
			\draw [thick, red] (3.2,0)  arc (0:120:0.2)node[anchor=south west]{$\alpha$};
			\draw [thick, green] (1.15,3) arc (0:250:0.2)node[anchor=south east]{$\beta$};
	\end{tikzpicture}}
	\hspace{4em}
	\subfloat[IO case]{%
		\begin{tikzpicture}
			\draw[thin,gray!40] (0,4) grid (0,4);
			\draw[<->] (0,0)--(4,0) node[right]{Re$(m_{\beta\beta})$};
			\draw[<->] (0,0)--(0,4) node[above]{Im$(m_{\beta\beta})$};
			\draw[line width=2pt,blue,-stealth](0,0)--(2,0) node[anchor=north]{$\boldsymbol{A}$};
			\draw[line width=2pt,red,-stealth](2,0)--(3,1.5) node[anchor=north west]{$\boldsymbol{B}$};
			\draw[line width=2pt,green,-stealth](3,1.5)--(2.5,2.2) node[anchor=north east]{$\boldsymbol{C}$};
			\draw [thick, red] (2.2,0)  arc (0:60:0.2)node[anchor=west]{$\alpha$};
			\draw [thick, green] (3.2,1.5)  arc (0:120:0.2)node[anchor=south west]{$\beta$};
	\end{tikzpicture}}
	\caption{Visualization of $m_{\beta\beta}$ as the sum of three complex numbers in examples of NO (left) and IO (right) scenarios. The NO example depicts a case where $m_{\beta\beta} = 0$.}
	\label{fig:mbbsums}
\end{figure}

The key distinguishing feature of the NO case is its funnel, an interval $[m_l',m_l'']$ within which it is possible to choose $\alpha$ and $\beta$ such that $m_{\beta\beta} = 0$ is attained. As shown in Eq.~\eqref{eq:mbb}, we can think of $m_{\beta\beta}$ as the sum of three complex numbers $A + Be^{i\alpha} + Ce^{i\beta}$, where $A\approx 0.67 m_1$, $B\approx 0.30 m_2$ and $C\approx 0.02 m_3$. This is depicted in Fig.~\ref{fig:mbbsums}, and in the IO case $m_3$ is too small for the sum to reach 0. In the NO case, many such closed triangles are found. Funnel edges $m_l'$ and $m_l''$ occur at $(\alpha,\beta) = (\pi,0)$ and $(\pi,\pi)$, respectively, and are analytic over the neutrino masses and mixing angles. Using best-fit values from NuFit v4.0 + SK \cite{Esteban2019, Nufit2018}, $m_l' = 2.330$~meV and $m_l'' = 6.535$~meV. For any $m_l \in [m_l',m_l'']$, there is a unique Majorana phase pair $(\alpha,\beta)$ which satisfies $m_{\beta\beta} = 0$. This choice of phase pair is extremely fine-tuned, leading to a statistical inaccessibility of the funnel; within the funnel interval in $m_l$, the fraction of $(\alpha,\beta)$ parameter space with $|m_{\beta\beta}| < 10^{-3}, 10^{-4}$ and $10^{-5}$~eV is $\approx 10^{-1}, 10^{-3}$ and $10^{-5}$, respectively.

The high-density banding and the inaccessibility of the funnel in the $|m_{\beta\beta}|-m_l$ parameter space due to the variation of the Majorana phases are a consequence of the parametrization chosen. In Bayesian language we may also say that the corresponding flat prior on the phases is an arbitrary choice given our lack of knowledge on the phases. Moreover, the two Majorana phases are somewhat redundant and what truly matters is the range of values in $|m_{\beta\beta}|$ that can occur for a given $m_l$. For phase angles and parameters derived from phase angles, a flat prior is the standard choice \cite{Agostini2017} and may be most easily described by a linear parametrization. To capture the effects of the Majorana phases we therefore introduce an effective phase parameter $0 \leq \Phi \leq 1$ which interpolates linearly in $|m_{\beta\beta}|$ between the boundaries of this permissible region. From Eq.~\eqref{eq:mbb}, these boundaries occur for pairs of $0$ and $\pi$ Majorana phases, which leads to the definitions
\begin{equation}
\label{eq:phieffNO}
	|m_{\beta\beta}^\text{NO}| = 
    \begin{cases}
        B + (A+C)(2\Phi - 1), & m_l \leq m_l', \\
        (A+B+C)\Phi,          & m_l' < m_l \leq m_l'', \\
        A + (B+C)(2\Phi - 1), & m_l > m_l'',
    \end{cases},
\end{equation}
for the NO case and
\begin{equation}
\label{eq:phieffIO}
	|m_{\beta\beta}^\text{IO}| = A + (B+C)(2\Phi - 1),
\end{equation}
in the IO case, where $A,B,C$ depend explicitly upon $m_l$. In this parametrization, $|m_{\beta\beta}|$ changes linearly as $\Phi$ is varied linearly. This includes the funnel region of the NO case.

\subsection{Bayesian Methodology}

A model $(H, \theta)$ is defined by both selecting a theory $H$ and a vector of values $\theta$ in the space of continuous parameters $\Theta_H$ spanning that theory. The following probability densities may be defined.
\begin{itemize}
	\item $\pi(\theta) \equiv P(H, \theta)$ is the prior belief in the model $(H, \theta)$ before data collection;
	\item $L_x(\theta) \equiv P(x|H, \theta)$ is the likelihood of observing data $x$ if the parametrized model $(H,\theta)$ is assumed to be true;
	\item $p(\theta|x) \equiv P(H, \theta|x)$ is the posterior belief in the model $(H,\theta)$ given observed data $x$.
\end{itemize}
While the posterior and prior distributions are probability densities over $\Theta_H$, and therefore normalisable on this domain, the likelihood is instead a probability density over the space of measurable data $X$. Bayes' Theorem relates these quantities in a manner analogous to statistical mechanics, where the posterior gives the probability density for the model to `occupy' state $\theta$ out of all possible parameter choices,
\begin{equation}
	p(\theta|x) 
	= \frac{L_x(\theta) \pi(\theta)}{\int_{\Theta_H} d\theta' L_{x}(\theta') \pi(\theta')}
	\equiv \frac{L_x(\theta) \pi(\theta)}{M_x^H},
	\label{bayesthm}
\end{equation}
where the normalisation factor $M_x^H$ is known as the marginal likelihood. In practice, the prior $\pi(\theta)$ is an educated guess, perhaps taking preceding experimental information into account, but presumed to be incomplete. As measured data becomes available, the prior probability is updated according to Bayes' Theorem, and each calculated posterior probability becomes the new prior. Given enough data, this process converges to the true best-fit model regardless of error in the prior.\footnote{An exception are priors using Dirac $\delta$-function shapes imprinting on the posterior, $p(\theta|x) = \pi(\theta) = \delta(\theta - \hat\theta)$, thereby leading to an incorrect result regardless of data. Similarly, sharp prior distributions tend to converge very slowly.}

In this paper, we take $0\nu\beta\beta$ decay to be mediated by light neutrino exchange as described above. Our model parameter space $\Theta_H$ is represented by the lightest neutrino mass $m_l$ and the effective Majorana phase parameter $\Phi$, $(m_l,\Phi) \in \Theta_H$. For simplicity, the other neutrino oscillation parameters are fixed at their best fit values and the mass ordering scenario, NO or IO, is considered to be known. The natural hypothesis to consider is the  observation of a certain number of signal events $n$ in a $0\nu\beta\beta$ experiment and so our data space $X$ is represented by the possible counts $n = 0, 1, 2,\dots$. This framework is easily applied to the comparison of multiple hypotheses, e.g. NO vs IO neutrino mass hierarchies, by computing marginal likelihoods for both models given the same observed data. The ratio of these quantities $K = M_x^{NO} / M_x^{IO}$ is the Bayes' factor of the NO hierarchy versus the IO hierarchy.

Data available from current or upcoming $0\nu\beta\beta$ decay experiments in isolation is insufficient to claim convergence of the posterior distribution -- instead predictions and bounds on model parameters are sought. All such quantities are expressible as posterior integrals \cite{Speagle2020}, which we calculate using samples obtained by MCMC with the Metropolis-Hastings algorithm~\cite{Brooks2011}.

%% file: lip.tex
We now consider the impact of prior selection on Bayesian inference from $0\nu\beta\beta$ decay. In the limit of perfectly precise measurements of observables which fully cover the given parameter space, the bias introduced by a prior vanishes. Unfortunately, this is not the situation for any real experimental outcome, and any assumption implicitly made by a prior distribution must contribute to the posterior, as demonstrated for $0\nu\beta\beta$ in particular by \cite{Gariazzo2018}. It is therefore advantageous, to the practitioner who wishes to avoid inferential bias, to choose priors which assume the least about the outcome of the experiment, and are in this sense ``uninformative''. In this section we study least-informative priors (LIPs), which are intended to maximise the expected information gained through measurement and inference, and develop a methodology for the case at hand.  
\subsection{Theoretical Construction of Reference Priors}
It is of import to first cite \cite{Dickey1973} for a justification of the overall methodology of this paper; that is, to conduct Bayesian inference with a diverse set of priors in order to arrive at the fullest picture of what our data implies. The situation in scientific experiment is not so different from that in psychological studies of human behaviour, where personal knowledge or opinions play a non-trivial role in even the most rational decision-making. By employing priors which take account of some full range of acceptable prior beliefs which an experimenter might hold, we can both gain confidence in inferences which hold broadly across the considered priors, and quantify the variation of inferred bounds or measurements between priors.

However, even if we are persuaded that no single prior can offer a complete understanding of any statistical inference, it becomes necessary to establish a \textit{reference prior} against which the performance of all other priors might be consistently compared. A general procedure is developed in \cite{Bernardo1979}, which may be applied to any inference, for identifying such a prior as the solution to an optimisation problem over information-gain; an LIP. This procedure is summarised as follows, applied separately for each parameter $\theta_i$ with all others fixed.

First, taking only the requirement that information gained from multiple measurements is additive, the information contained in a distribution $P(\theta_i)$ is~\cite{Shannon1948, Lindley1956}
\begin{align}
	I = \int_{P(\theta_i)\neq 0}d\theta_i P(\theta_i) \log P(\theta_i),
\end{align}
which is familiar in physics as the Boltzmann-Gibbs entropy for a continuous collection of states, up to a constant factor \cite{Jaynes1965}. We then take an experiment $E$, which measures data $x$, with likelihood function $L_x(\theta_i)$. Following~ \cite{Lindley1956}, Ref.~\cite{Heavens2018} calculates the expected information gain of prior $\pi(\theta_i)$ as
\begin{align}
\label{eq:distinfo}
	I\{E,\pi\} 
	&= \int dx \int d\theta_i L_x(\theta_i)\pi(\theta_i) 
	   \log\left[\frac{p(\theta_i|x)}{\pi(\theta_i)}\right] \nonumber\\
    &= \int dx\, M_x\! \int d\theta_i p(\theta_i|x) 
       \log\left[\frac{p(\theta_i|x)}{\pi(\theta_i)}\right]
\end{align}
where $p(\theta_i|x)$ is the posterior given by Bayes' Theorem, and $M_x = \int d\theta_i L_x(\theta_i) \pi(\theta_i)$ is the marginal likelihood of data $x$. The inner integral on the second line of Eq.~\eqref{eq:distinfo} is known as the Kullback-Leibler divergence $K[p,\pi]$ between $p(\theta_i|x)$ and $\pi(\theta_i)$ \cite{Heavens2018}, of which $I\{E,\pi\}$ is therefore an expectation value over the data-space. Whether phrased as a prior or posterior integral, this quantity depends strongly on the choice of prior, which appears implicitly in the inference of the posterior and as the measure over $\theta_i$ in the marginal likelihood. Letting $E(k)$ indicate $k$ independent replications of experiment $E$, the quantity $I\{E(\infty),\pi\}$ describes the vagueness of prior $\pi(\theta_i)$ \cite{Bernardo1979}, as an infinite quantity of well-defined experiments must arrive at the same precise measurement, and so a greater expected information-gain through inference implies that more information was missing to begin with.

The prior $\pi$ which maximizes $I\{E(\infty),p\}$ cannot simply be selected because an infinite quantity of information is needed to measure any parameter exactly. Instead, a limit must be taken as the number of measurement repetitions $k$ approaches $\infty$. For a given $k$, the reference prior $\pi_k(\theta)$ is defined as that among all permissible priors \cite{Berger2009} which maximizes $I\{E(k),\pi\}$, where permissibility is defined using boundedness and consistency arguments over compact subsets of the parameter space. Given any measurement $x$, the reference posterior $p_k(\theta_i|x)$ corresponding to prior $\pi_k(\theta)$ is calculable by Bayes' Theorem, and assuming compactness on the set of possible posteriors, the limit $p(\theta_i|x) = \lim_{k\to\infty} p_k(\theta_i|x)$ is well-defined. Due to consistent validity of Bayes' Theorem across the measurement domain, a prior $\pi(\theta_i)$ proportional to $p(\theta_i|x)/L_x(\theta_i)$ is then a well-defined LIP which is independent of $x$.

Obtaining the LIP via a posterior limit of course does not feel very efficient, but so long as certain regularity conditions are met \cite{Berger2009}, a limiting sequence among priors which still maximizes Eq.~\eqref{eq:distinfo} may be found:
\begin{align}
\label{eq:lipk}
    \pi_k(\theta_i) \propto \exp\left\{\int dx L_x(\theta_i) \log\left[p^*(\theta_i|x^k)\right]\right\}.
\end{align}
Here, $x^k$ is a collection of data from $k$ repeated measurements, $\pi^*(\theta)$ is an initialization prior chosen among any in the permissible set, and $k$ is taken to be large enough that the posterior $p^*(\theta|x^k)$ induced by prior $\pi^*(\theta)$ is dominated by the characteristics of the likelihood rather than by that prior.

A subtlety of this construction is that a direct limit $\lim_{k\to\infty} \pi_k$ can be poorly behaved at singularities and boundaries, resulting in a comb-like LIP. In such cases, the LIP can instead be defined as a conditioned limit at some well-behaved parameter point $\theta_{i,0}$ \cite{Berger2009},
\begin{equation}
\label{eq:liplimit}
    \pi(\theta_i) = 
    \lim_{k\to\infty} \frac{\pi_k(\theta_i)}{\pi_k(\theta_{i,0})}.
\end{equation}
In addition to maximizing the expected inferential information-gain, the generated LIP enjoys simple Jacobian transformation under re-parametrizations of $\theta$ \cite{Bernardo1979}. 

\subsection{Implementation of LIP Algorithm for $0\nu\beta\beta$}
The application of LIPs to neutrino oscillation experiments is explored in \cite{Heavens2018}. When an experiment is such that the asymptotic posterior is well-approximated by a Gaussian distribution (a condition known as asymptotic posterior normality), it can be shown that the LIP is simply the multivariate Jeffreys prior of the likelihood \cite{Bernardo1979}. This assumption holds and significantly simplifies computation for oscillation studies of the neutrino mass splittings, and cosmological studies of the sum-of-masses $\Sigma$. However, in $0\nu\beta\beta$ decay, the observable of interest $|m_{\beta\beta}|$ may be asymptotically small, or may even vanish if neutrinos are not Majorana particles. Asymptotic posterior normality can therefore not be said to hold for any $0\nu\beta\beta$ decay search, and instead we follow the computational procedure set out by Berger \cite{Berger2009} for solving Eqs.~\eqref{eq:lipk} and \eqref{eq:liplimit} in full generality.

So long as an amenable likelihood model is chosen, the LIP computation expressed in Eq.~\eqref{eq:lipk} is well-suited to a sampling procedure. We employ Berger's method of selecting $m$ sets of $k$ likelihood samples each: for each set, Bayesian inference is made using a fiducial flat prior and the product of the $k$ likelihoods, as the samples are treated as occurring from independent experiments. The resultant posterior distribution is used to compute Eq.~\eqref{eq:lipk}, where the numerical integration is performed by averaging over the $m$ sample sets.

Following \cite{Caldwell2017}, we define our $0\nu\beta\beta$ measurement model to be a Poisson counting experiment, where the likelihood of observing $n$ counts given a background expectation $\lambda$ and signal expectation $\nu$ is
\begin{equation}
    L_n(\nu) = e^{-\lambda-\nu}\frac{(\lambda+\nu)^n}{n!}.
\end{equation} 
The number of expected signal events $\nu$ is related to the $0\nu\beta\beta$ decay half-life by
\begin{equation}
    \nu = \frac{\mathcal{E}N_A\log 2}{m_\text{iso}} T^{-1}_{1/2},
\end{equation}
where $N_A$ is Avogadro's constant, $m_\text{iso}$ is the molar mass of the enriched isotope used in detection and $\mathcal{E}$ is the sensitive exposure, also accounting for detection efficiency. In our simulations for the LEGEND-200 $^{76}$Ge experiment, we take one year of runtime with $m_\text{iso} = 75.921$~u and $\lambda = 1.7\cdot 10^{-3}$~cts/(kg$\cdot$yr)~$\cdot$~$\mathcal{E}$ with sensitive exposure $\mathcal{E} = 119$~kg$\cdot$yr \cite{Agostini2017}. The half-life $T_{1/2}(|m_{\beta\beta}|)$ also depends on the isotope through the phase space factor and nuclear matrix element, where we use the values $G_{0\nu} = 3.04\cdot 10^{-26}$~yr$^{-1}$eV$^{-2}$ and $|\mathcal{M}_{0\nu}| = 4.32$ \cite{Faessler2008}.

Berger's algorithm discussed above applies only to a single parameter $\theta_i$, and therefore in a multi-parameter problem such as ours, the LIP must be obtained sequentially. Following \cite{Heavens2018}, at step $j$ in the iteration, the prior on $\theta_j$ is computed using fixed values of all $\theta_{i>j}$, written $\pi(\theta_j | \theta_{i>j})$. The likelihood function is then marginalised by parameter $\theta_j$,
\begin{equation}
    L_x(\theta_{i > j}) = \int d\theta_j \pi(\theta_j | \theta_{i>j}) L_x(\theta_{i \geq j}).
\end{equation}
Note that this procedure must be repeated for each combination of parameter values for which we seek to know the LIP, placing a strong bottleneck on achievable precision. At the end of the iteration, the total LIP is given by the product
\begin{equation}
    \pi(\theta) = \prod_{j\leq n} \pi(\theta_j | \theta_{i>j}).
\end{equation}
For a non-separable likelihood function, this depends on the ordering of parameters \cite{Heavens2018}, with more impactful parameters customarily ordered first; we therefore take $\theta_1 \equiv m_l$ and $\theta_2 \equiv \Phi$. The resultant two-stage LIP algorithm is summarized in Fig.~\ref{fig:LIPflowchart} in Appendix~\ref{sec:flowchart}. A ``free-phi'' approximation is also considered, in which the above multi-parameter procedure is still followed, but the $\Phi$-likelihood $L_n^{\Phi}(\Phi)$ is computed over a flat $m_l$ prior, thereby removing costly interpolation evaluations.

From a practical standpoint, the number of repetitions $m$ impacts the precision of the final LIP, while the sample quantity $k$ affects its accuracy. If $m$ is too small, the prior may be noisy but still accurate, while a $k$ far from the convergence region could lead to a prior which is far from least-informative. We select $m = 100$ for our simulations, sufficiently large to be near-convergence, but small enough to avoid precision errors as the product of likelihoods dips near $10^{-200}$. The outer loops of the algorithm are parallelizable, and so a 16-core MPI implementation of the algorithm leads to significant speed-up, allowing for $k$ up to $2000$ with $\sim 12$-hour run-times.

%% file: results.tex
\subsection{Generated LIPs for LEGEND-200}
To illustrate the above procedure, we choose the future $0\nu\beta\beta$ decay experiment LEGEND-200 \cite{Myslik2018} as the basis for a measurement example. LEGEND-200 plans to use 175~kg of the isotope $^{76}$Ge to achieve a $3\sigma$ sensitivity to half-lives greater than $10^{27}$~yr. This corresponds to an expected number of background events of $\lambda = 1.7\cdot 10^{-3}$~cts/(kg$\cdot$yr)~$\cdot$~$\mathcal{E}$ with sensitive exposure $\mathcal{E} = 119$~kg$\cdot$yr \cite{Agostini2017} taken over one year of runtime as an example to illustrate our algorithm. Here the sensitive exposure $\mathcal{E}$ is defined as the product of the total exposure with fiducial volume and signal detection efficiencies, also accounting for a $2\sigma$ region of interest around the decay energy. Two configurations of the LIP algorithm were considered: the full two-parameter integration discussed above and specified in Figure~\ref{fig:LIPflowchart} in the Appendix, and the free-phi approximation.

\begin{figure}[t!]
	\centering
	\includegraphics[width=\textwidth]{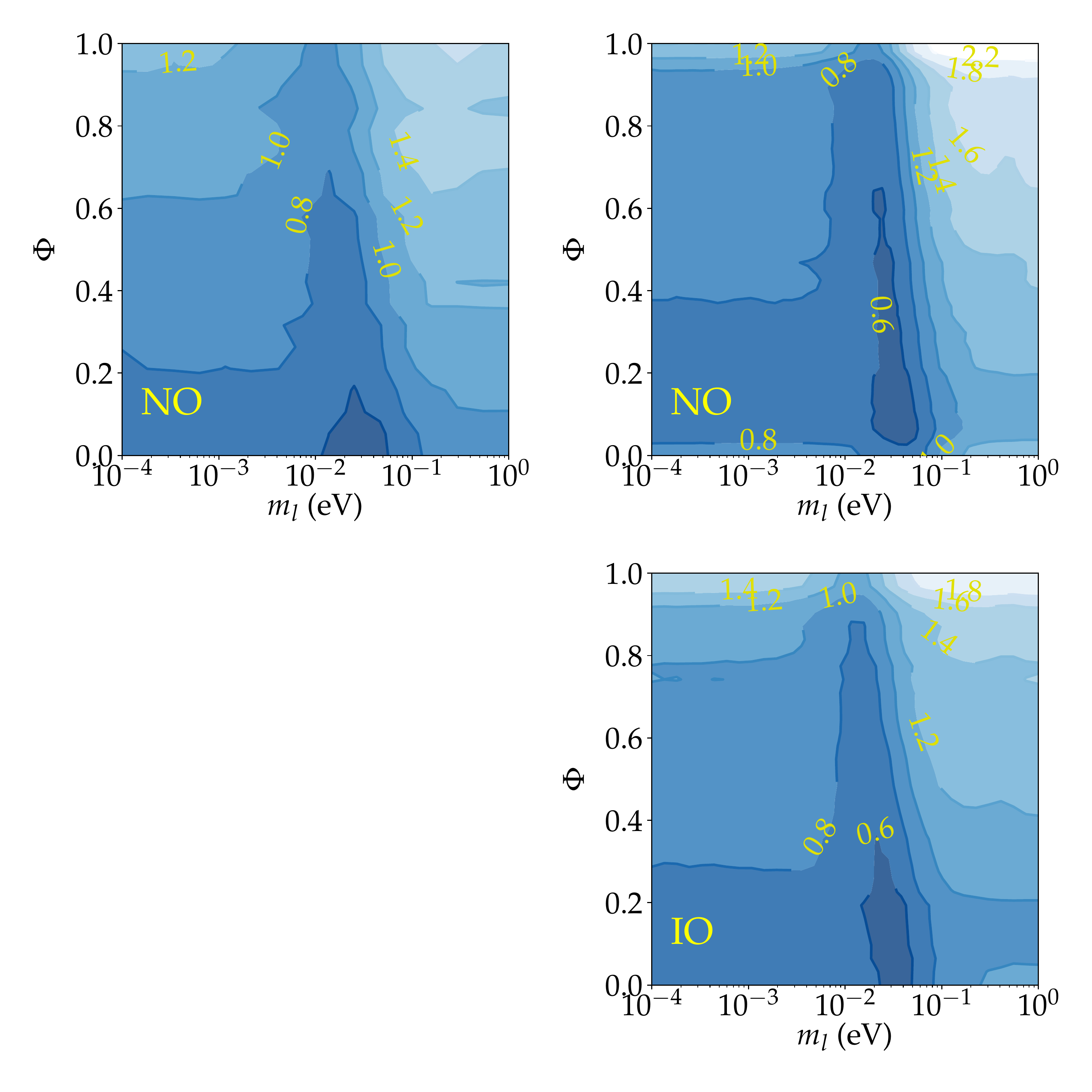}
	\caption{Least-Informative Priors $\pi(m_l,\Phi)$ in terms of the lightest neutrino mass $m_l$ and the effective Majorana phase parameter $\Phi$ based on a likelihood using the LEGEND-200 experiment (expected background events $\lambda = 1.7\cdot 10^{-3}$~cts/(kg$\cdot$yr)~$\cdot$~$\mathcal{E}$ with sensitive exposure $\mathcal{E} = 119$~kg$\cdot$yr). The LIPs in the NO scenario are computed using the full algorithm (top left) and free-phi integration (top right). The bottom right plot shows the LIP for IO using free-phi integration.}
	\label{fig:priors}
\end{figure}
The results of our simulation, with $2\times 10^5$ likelihood draws at each parameter point $(m_l, \Phi)$ across a grid with resolutions $\Delta \Phi = 0.1$ and $\Delta \log_{10}(m_l) = 0.2$, are shown in Fig.~\ref{fig:priors}. The top left and top right plots show the LIP $\pi(m_l,\Phi)$ in the NO scenario calculated using the full simulation and the free-phi approach, respectively. They demonstrate that the free-phi approximation is generally valid throughout the parameter space. The greatest deviation occurs for large values of both parameters, a region where relevant likelihoods tend to be very low, and the impact upon posterior inference is therefore expected to be negligible. It should be noted that slow convergence at parameter boundaries causes erroneous growth of the raw LIP, producing boundary walls which are locally smoothed in post-processing to produce the visuals throughout this work, and before MCMC or information computations are performed. Signal-to-noise ratio (SNR) analysis of repeated trials showed a minimum SNR of $1000$, or $30$~dB, between individual generated LIPs and averages of three LIPs, as a reference. Practically, this corresponds to a noise amplitude of at most $\pm 0.05$ everywhere in the LIP distribution.

In the right column of Fig.~\ref{fig:priors}, LIPs computed for NO (top) and IO (bottom) neutrino mass orderings are compared and seen to be structurally similar. Both priors feature a near-linear increase with $\Phi$, and a significant trough in $m_l$ between $10^{-2}$ and $10^{-1}$~eV. On either side of this trough, the prior is nearly flat in $m_l$, with higher density for $m_l > 0.1$~eV. Note that these functions are distributions over $m_l$, plotted on a logarithmic scale, rather than distributions over $\log(m_l)$, which would feature the presence of an additional factor $1/m_l$ from the Jacobian transformation.

It is significant that a predominantly flat prior in $m_l$ emerges from a first-principles Bayesian simulation, perhaps indicating the naturalness of a flat prior for unknown particle masses. The trough may be understood physically as the region of parameter space where LEGEND-200 has the greatest propensity to make either a measurement or an exclusion; the LIP therefore reduces the weight in this region so that any inferences made can be said to more fully data-driven. 

\subsection{Comparison of Inferred Bounds on $m_l$}
We evaluate the performance of the computed LIPs by utilizing them (after bi-cubic spline interpolation) in our MCMC analysis, equipped with the effective parametrisation $(m_l, \Phi)$. Only the experimental likelihood from LEGEND-200 is included, modelled with a Poisson distribution following \cite{Caldwell2017} with $\lambda = 1.7\cdot 10^{-3}$~cts/(kg$\cdot$yr)~$\cdot$~$\mathcal{E}$ with sensitive exposure $\mathcal{E} = 119$~kg$\cdot$yr \cite{Agostini2017}.

\begin{figure}[t!]
	\centering
	\includegraphics[width=\textwidth]{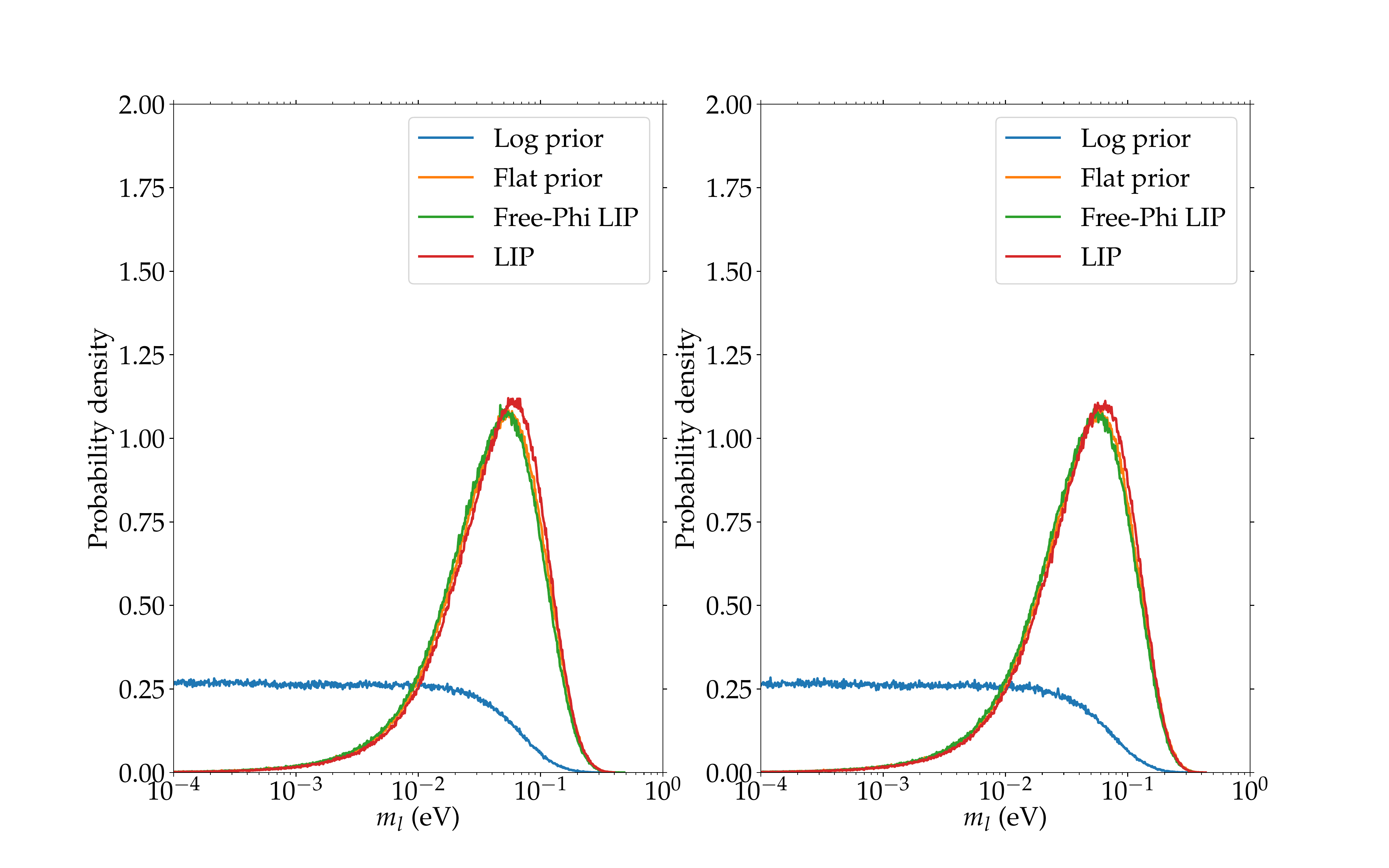}
	\caption{Posterior distributions over $m_l$, computed from marginalised MCMC samples for the projected LEGEND-200 likelihood for NO (left) and IO (right) in the case of non-observation (count $n=0$). The curves correspond to the different priors used: Log prior (log-flat $m_l$ and flat $\Phi$, blue), Flat prior (flat $m_l$ and flat $\Phi$, orange), Free-Phi LIP (LEGEND-200 LIP with free-phi integration, green) and LIP (full LEGEND-200 LIP, red).} 
	\label{fig:n0posteriors}
\end{figure}
We consider first the case where LEGEND-200 registers $n = 0$ signal events during its run. We include $10^7$ MCMC samples, using a Gaussian proposal distribution of width $10^{-2}$ in both $m_l$ and $\Phi$. The resulting posterior distributions for flat (in both $m_l$ and $\Phi$) and log-flat (flat in $\log m_l$ and $\Phi$) priors as well as both LIP implementations are shown in Fig.~\ref{fig:n0posteriors} for NO (left) and IO (right). A further confirmation of the strong match between free-phi and full LIP calculations is gained, and both are seen to lead to very similar inferences as the flat prior. In the NO case, these three priors exclude $m_l > 0.11$~eV at a $90\%$ credibility level \footnote{Bayesian credibility intervals may be thought of as analagous to frequentist confidence intervals.}, while the log-flat prior excludes $m_l > 0.03$~eV; in the IO case, these upper bounds are expectedly slightly higher, though within simulation error.

\begin{figure}[t!]
	\centering
	\includegraphics[width=\textwidth]{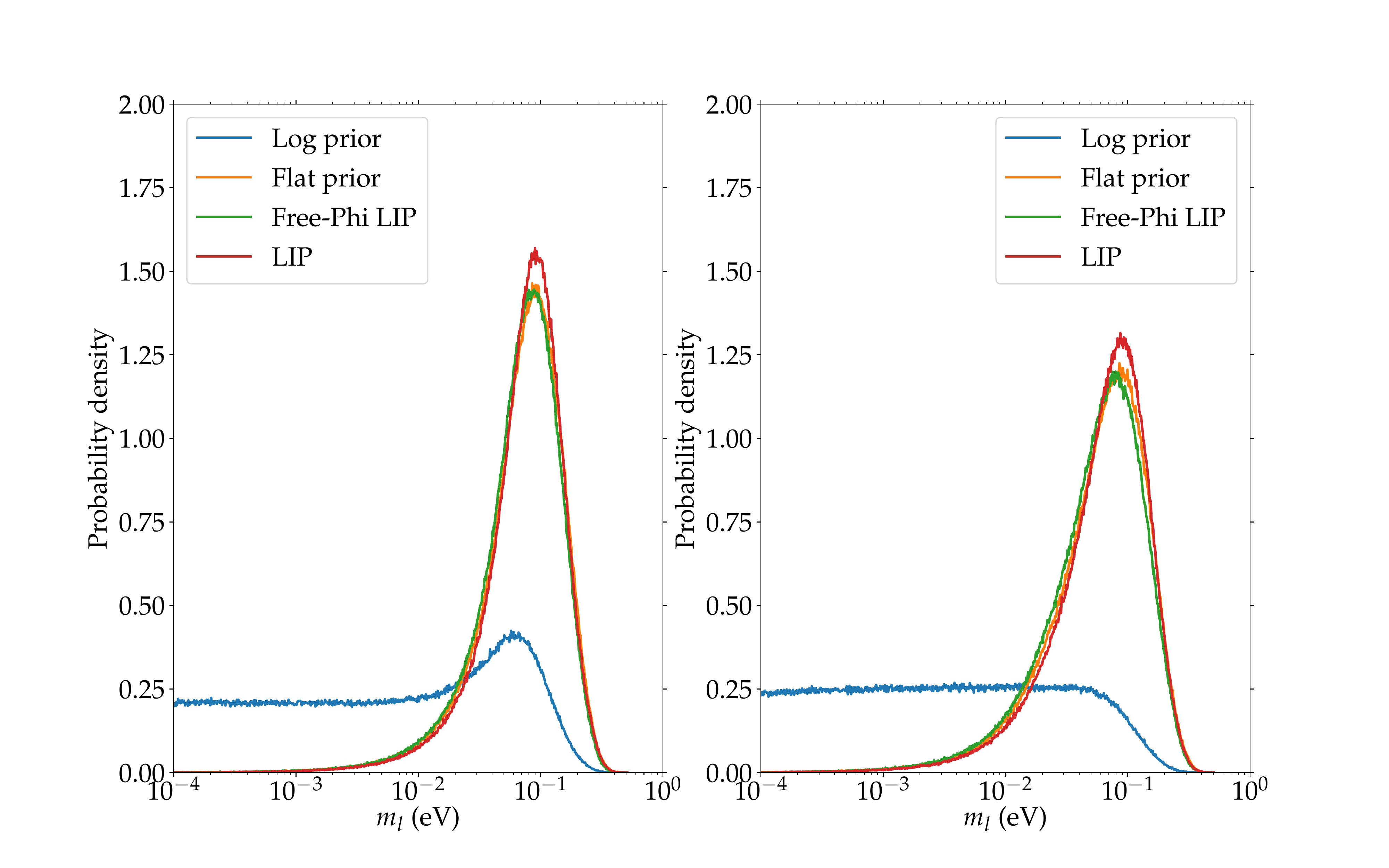}
	\caption{As Fig.~\ref{fig:n0posteriors}, but in the case of observation of one signal event, $n = 1$.} 
	\label{fig:n1posteriors}
\end{figure}
Next we set the observed count to $n = 1$, and repeat the inference procedure, with results shown in Fig.~\ref{fig:n1posteriors}. Again both calculated LIPs give similar posteriors to the flat prior. In the NO case, Fig.~\ref{fig:n1posteriors}~(left), these three priors result in $m_l = 90\pm 50$~meV. In the IO case, Fig.~\ref{fig:n1posteriors}~(right), the flat and LIP measurements give slightly lower values, but with comparable precision. However, the log-flat prior fails to make any measurement in either hierarchy, instead placing an exclusion at $90\%$ credibility on $m_l > 75$~meV in the NO case, and on $m_l > 45$~meV in the IO case, as the measurement falls squarely within the region extending to low $m_l$ which log-flat priors are intended to probe. It is not surprising that LEGEND-200 fails to distinguish between the neutrino-mass hierarchies, as it does not probe sufficiently small $|m_{\beta\beta}|$ where the allowed parameter regions notably diverge.

\subsection{Information Content of Inferences}

\begin{table}[t!]
	\centering
	\begin{tabular}{c|cccc}  
		\hline
		Prior        &  NO, $n=0$ &  IO, $n=0$ &  NO, $n=1$ &  IO, $n=1$ \\
		\hline
		Log prior    &     14.64  &     14.63  &     15.90 &     15.23  \\
		Flat prior   &     20.63  &     20.61  &     20.63  &     20.73  \\  
		Free-Phi LIP &     21.92  &     21.94  &     21.76  &    21.81  \\ 
		LIP          & \bf{22.08} & \bf{22.09} & \bf{21.90} & \bf{22.04} \\
		\hline
	\end{tabular}
	\caption{Kullback-Leibler divergences (in bits) for different priors, computed from MCMC LEGEND-200 posteriors in the case of NO and IO as well as observed counts $n = 0, 1$, as indicated. The entries in bold denote the highest divergence achieved among the different priors used.} 
	\label{tab:kldivergences}
\end{table}
However, the value of LIPs does not lie in their propensity to give a conservative bound or measurement (which the flat prior already achieves), but in their trustworthiness as a reference prior with minimised bias. In Table~\ref{tab:kldivergences}, we report the Kullback-Leibler divergences of each MCMC posterior against its prior, using the same LEGEND-200 likelihood as above. Error propagation was performed by considering a noisy LIP of the form $\pi(\theta) = \pi_{true}(\theta) \pm n_\pi(\theta)$, which in the limit of small noise (and assuming that variation in the posterior is dominated by variation in the prior) corresponds to noisy Kullback-Leibler divergence:
\begin{equation}
K[p,\pi] = K[p,\pi_{true}] \pm \frac{1}{\log(2)} \int d\theta p(\theta) \frac{n_\pi(\theta)}{\pi_{true}(\theta)} 
\end{equation}
where the integrated error may be interpreted as the posterior expectation value of the inverse of the signal-to-noise ratio distribution for $\pi(\theta)$. Computation of this integral for the worst-case noise distribution mentioned in Section 4.1 gave an error of $\pm 0.03$~bits, leading us to quote our divergence values to $0.01$~bit precision.

The results give a numerical confirmation of the similarity between free-phi and full LIP computations, whose divergences consistently fall within $0.25$~bits. In all configurations, the LIPs outperform both standard priors in information gain, as expected from their construction.

%% file: conclusion.tex
Bayesian parameter inference is a common tool to constrain or determine parameters in particle physics. An inherent issue in this context is the choice of a prior distribution over the model parameters. We have here focussed on the neutrino parameter space relevant to $0\nu\beta\beta$ decay searches, specifically the lightest neutrino mass $m_l$ and an effective Majorana phase parameter $\Phi$ encapsulating the effect of the Majorana phases in the lepton mixing matrix. Given that $0\nu\beta\beta$ decay has not been observed yet, prior distributions are expected to have a strong impact on the conclusions drawn.

We have adapted an algorithm for computing least-informative priors for a given experiment via likelihood-sampling to the case of $0\nu\beta\beta$ direct searches, resulting in exact and approximate parallelised implementations. The LIPs were seen to take the form of a flat-$m_l$, linear-$\Phi$ distribution broken by a trough between $m_l = 10^{-2}$ and $10^{-1}$~eV. We demonstrated that for the proposed 200~kg $^{76}$Ge LEGEND experiment, these priors give similar posterior bounds to the usually adopted flat prior for both neutrino orderings, and in both observation and non-observation scenarios. Furthermore, the LIPs were seen in nearly all cases to outperform both standard flat and logarithmic priors as far as their information-gain during MCMC inference is concerned. This supports the functionality of the adapted algorithm and strengthens the argument for the use of LIPs as reference priors for $0\nu\beta\beta$ decay searches.

Natural extensions of this work include a study of the variation in LIP performance across proposed experiments of diverse background levels and exposures, simulation of LIPs for the usual parametrization using two Majorana phases and research towards the construction of a prior which is jointly least-informative over both the $0\nu\beta\beta$ observable $|m_{\beta\beta}|$ and the cosmology observable $\sum m_i$.

%% file: flowchart.tex
%
\begin{figure}[h!]
	\centering
	\begin{tikzpicture}[node distance = 1em, auto]
		\tikzstyle{every node}=[font=\small]
		\node [label] (loopphi) {For each $\Phi^*$ in range $[\Phi]$};
		\node [label, below= of loopphi] (loopml) {For each $m_l^*$ in range $[m_l]$};
		
		\node [label, below= of loopml] (loopj) {For $j = 1:m$};
		
		\node [block, below= of loopj] (sample) {Take $k$ likelihood samples $\{x_{1j},\dots,x_{kj}\}$ of signal $n$ from $L_n(m_l^*,\Phi^*)$};
		\node [block, below= of sample] (ml) {Calculate ML $c_j = \displaystyle{\int} \prod_{i=1}^k L_{x_{ij}}(m_l',\Phi^*) \pi^*(m_l') dm_l'$};
		\node [block, below= of ml] (logpost) {Calculate log-posterior $r_j(m_l) = \log\left[\prod_{i=1}^{k} L_{x_{ij}}(m_l^*,\Phi^*) \pi^*(m_l^*)/c_j\right]$};
		
		\begin{pgfonlayer}{back3}
			\node [bigbox, minimum width=30em, fill=blue!10, fit={(loopj) (sample) (ml) (logpost)}] (loopjbox) {};
		\end{pgfonlayer}
		
		\node [block, below= of loopjbox] (lipml) {Calculate LIP point $\pi(m_l^*|\Phi^*) = \exp\left[\frac{1}{m}\sum_{j=1}^{m}r_j(m_l^*)\right]$};
		
		\begin{pgfonlayer}{back2}    
			\node [bigbox, minimum width=36em, fill=blue!6, fit={(loopml) (loopjbox) (lipml)}] (loopmlbox) {};
		\end{pgfonlayer}
		
		\node [block, below= of loopmlbox] (interpml) {Interpolate over pairs $(m_l^*, \pi(m_l^*|\Phi^*))$ to get smooth $\pi(m_l | \Phi^*)$};
		\node [block, below= of interpml] (philike) {Define $\Phi$-likelihood: $L^\Phi_n(\Phi^*) = \displaystyle{\int} L_n(m_l',\Phi^*) \pi(m_l'|\Phi^*) dm_l'$};
		
		\node [label, below= of philike] (loopj2) {For $j = 1:m$};
		
		\node [block, below= of loopj2] (sample2) {Take $k$ likelihood samples $\{x_{1j},\dots,x_{kj}\}$ of signal $n$ from $L^\Phi_n(\Phi^*)$};
		\node [block, below= of sample2] (ml2) {Calculate ML $c_j = \displaystyle{\int} \prod_{i=1}^{k} L^\Phi_{x_{ij}}(\Phi') \pi^*(\Phi') d\Phi'$};
		\node [block, below= of ml2] (logpost2) {Calculate log-posterior $r_j(\Phi^*) = \log\left[\prod_{i=1}^k L^\Phi_{x_{ij}}(\Phi^*) \pi^*(\Phi^*)/c_j\right]$};
		
		\begin{pgfonlayer}{back2}
			\node [bigbox, minimum width=30em, fill=blue!6, fit={(loopj2) (sample2) (ml2) (logpost2)}] (loopjbox2) {};
		\end{pgfonlayer}
		
		\node [block, below= of loopjbox2] (lipphi) {Calculate LIP point $\pi(\Phi^*) = \exp\left[\frac{1}{m}\sum_{j=1}^{m}r_j(\Phi^*)\right]$};
		
		\begin{pgfonlayer}{back1} 
			\node [bigbox, minimum width=42em, fill=blue!3, fit={(loopphi) (loopmlbox) (interpml) (philike) (loopjbox2) (lipphi)}] (loopphibox) {};
		\end{pgfonlayer}
		
		\node [block, below= of loopphibox] (interpphi) {Interpolate 2D surface over points $(m_l^*,\Phi^*, \pi(m_l^*,\Phi^*) \equiv \pi(\Phi^*) \pi(m_l^* | \Phi^*))$ to get smooth $\pi(m_l, \Phi)$};
		
		\path [line] (sample) -- (ml);
		\path [line] (ml) -- (logpost);
		\path [line] (loopjbox) -- (lipml);
		\path [line] (loopmlbox) -- (interpml);
		\path [line] (philike) -- (loopjbox2);
		\path [line] (interpml) -- (philike);
		\path [line] (sample2) -- (ml2);
		\path [line] (ml2) -- (logpost2);
		\path [line] (loopjbox2) -- (lipphi);
		\path [line] (loopphibox) -- (interpphi);
		
	\end{tikzpicture}
	\caption{Algorithm for generating a least-informative prior on the $(m_l,\Phi)$ parameter space, given experimental settings $\lambda$, $m_\text{iso}$, and $\mathcal{E}$, and fiducial priors $\pi^*(m_l^*)$ and $\pi^*(\Phi^*)$.}
	\label{fig:LIPflowchart}
	\vspace{-12pt}
\end{figure}
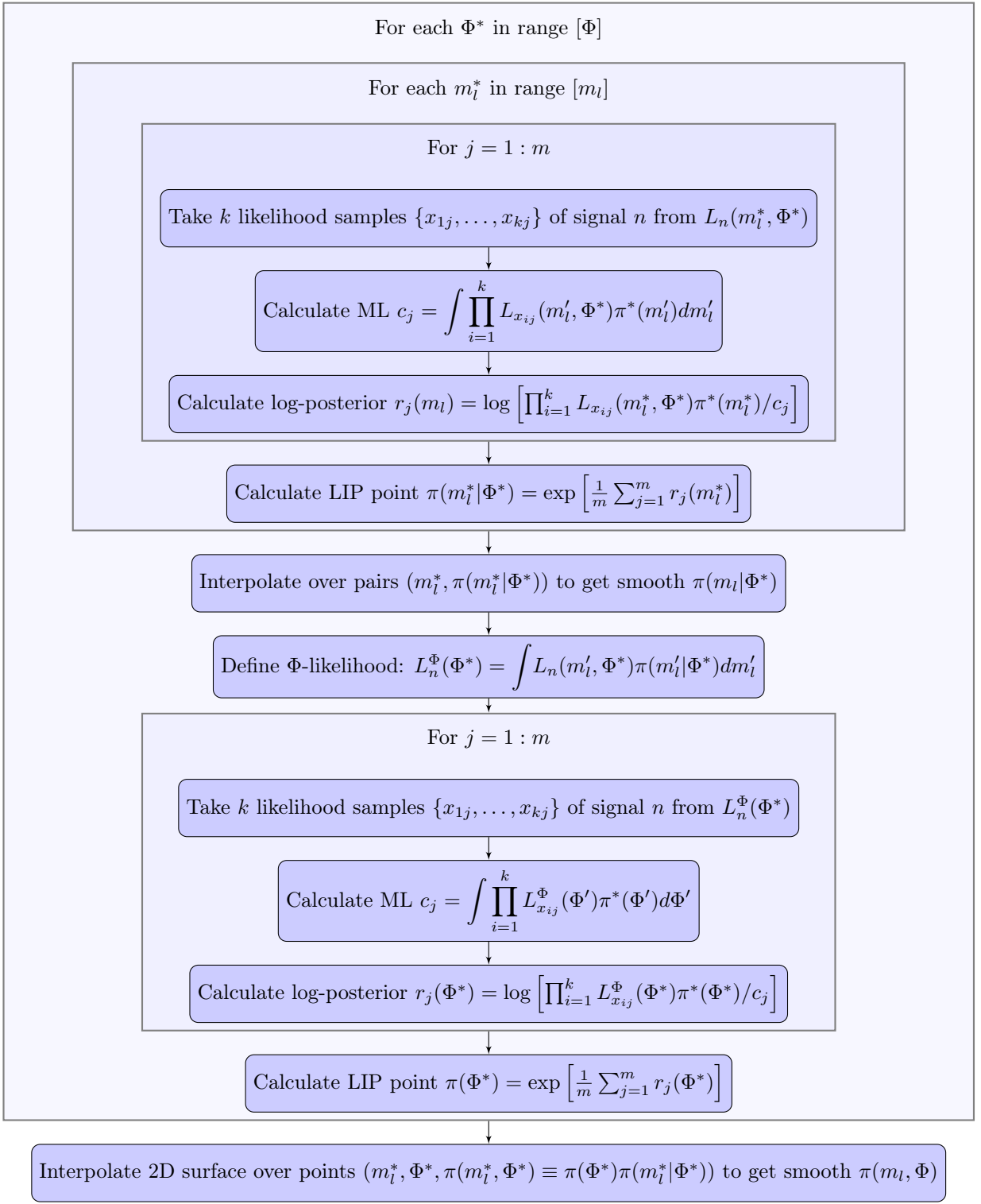